\documentclass[%
reprint,
 amsmath,amssymb,
 aps,
 prd,
]{revtex4-1}

\usepackage[utf8]{inputenc}
\usepackage{natbib}
\usepackage{dblfnote}
\DFNalwaysdouble
\usepackage{slashed}
\usepackage{graphicx}
\usepackage{dcolumn}
\usepackage{bm}
\usepackage{aas_macros}
\usepackage{color}

\newcommand{\mrm}[1]{_{\rm #1}}
\renewcommand{\d}{{\rm d}}
\newcommand{\citeeq}[1]{Eq.~(\ref{#1})}
\newcommand{\citefig}[1]{Fig.~\ref{#1}}

\begin{document}

\begin{flushright}
	CERN-TH-2019-084
\end{flushright}

\title{Constraining primordial black hole masses with the isotropic gamma ray background}

\author{Alexandre Arbey}
\altaffiliation[Also at ]{Institut Universitaire de France, 103 boulevard Saint-Michel, 75005 Paris, France.}%
\email{alexandre.arbey@ens-lyon.fr}
\affiliation{
	Univ Lyon, Univ Lyon 1, CNRS/IN2P3, Institut de Physique Nucl\'eaire de Lyon, UMR5822, F-69622 Villeurbanne, France
}

\author{J\'er\'emy Auffinger}
\email{j.auffinger@ipnl.in2p3.fr}
\affiliation{
	Institut d'Astrophysique de Paris, UMR 7095 CNRS, Sorbonne Universit\'es,
	98 bis, boulevard Arago, F-75014, Paris, France \\
	Univ Lyon, Univ Lyon 1, CNRS/IN2P3, Institut de Physique Nucl\'eaire de Lyon, UMR5822, F-69622 Villeurbanne, France \\
	D\'epartement de Physique, \'Ecole Normale Sup\'erieure de Lyon, F-69342 Lyon, France
}

\author{Joseph Silk}
\email{joseph.silk@physics.ox.ac.uk}
\affiliation{%
	Institut d'Astrophysique de Paris, UMR 7095 CNRS, 
	Sorbonne Universit\'es,
	98 bis, boulevard Arago, F-75014, Paris, France \\  The Johns Hopkins University, Department of Physics and Astronomy, Baltimore, Maryland 21218, USA  \\Beecroft Institute of Particle Astrophysics and Cosmology, University of Oxford, Oxford OX1 3RH, UK 
}%
\date{\today}

\begin{abstract}
Primordial black holes can represent all or most of the dark matter in the window $10^{17}-10^{22}\,$g. Here we present an extension of the constraints on PBHs of masses $10^{13}-10^{18}\,$g arising from  the isotropic  diffuse gamma ray background. Primordial black holes evaporate by emitting Hawking radiation that should not exceed the observed background. Generalizing from monochromatic distributions of Schwarzschild black holes to extended mass functions of Kerr rotating black holes, we show that the lower part of this mass window can be closed for near-extremal black holes.
\end{abstract}

\maketitle


\section{\label{sec:intro}Introduction}

Primordial Black Holes (PBHs) are the only candidate able to solve the Dark Matter (DM) issue without invoking new physics. Two mass windows are still open for the PBHs to contribute to all or most of the DM: the $10^{17} - 10^{19}\,$g range, recently re-opened by \cite{Katz2018} after revisiting the $\gamma$-ray femtolensing constraint, and the $10^{20}-10^{22}\,$g range \cite{Niikura2017}, from HST microlensing probes of M31. PBHs are believed to have formed during the post-inflationary era, and subsequently evolved through accretion, mergers and Hawking Radiation (HR). If the PBHs were sufficiently numerous, that is to say if they contribute to a large fraction of DM, HR from PBHs may be the source of observable background radiation.

In this Letter, we update the constraints  on the number density of PBHs by observations of the diffuse Isotropic Gamma Ray Background (IGRB) \cite{Carr2010}, taking into account the latest FERMI-LAT data and, as  new constraints, the  spin of PBHs and extension of the PBH mass function. Our assumption is that part of the IGRB comes from the time-stacked, redshifted HR produced by evaporating PBHs distributed isotropically in the extragalactic Universe. Those PBHs must have survived at least until the epoch of CMB transparency for the HR to be able to propagate in the intergalactic medium. This sets the lower boundary on the PBH mass $M\mrm{min} \approx 5\times10^{13}\,$g. Furthermore, the HR peaks at an energy which decreases when the PBH mass increases. This sets the upper boundary for the PBH mass $M\mrm{max} \approx 10^{18}\,$g as the IGRB emission does not constrain the photon flux below $100\,$keV.

This Letter is organized as follows: Section~\ref{sec:Hawking} gives a brief reminder of HR physics, Section~\ref{sec:IGRB} describes the IGRB flux computation and Section~\ref{sec:results} presents the new constraints obtained with Kerr and extended mass function PBHs.

\section{\label{sec:Hawking}Kerr PBH Hawking radiation}

Black Holes (BHs) emit radiation and particles similar to blackbody radiation \cite{Hawking1975} with a temperature linked to their mass $M$ and spin parameter $a \equiv J/M \in [0,M]$ ($J$ is the BH angular momentum) through
\begin{equation}
	T \equiv \dfrac{1}{2\pi}\left( \dfrac{r_+ - M}{r_+^2 + a^2} \right)\,, \label{eq:temperature}
\end{equation}
where $r_+ \equiv M + \sqrt{M^2-a^2}$ and we have chosen a natural system of units with $G = \hbar = k\mrm{B} = c = 1$. The number of particles $N_i$ emitted per units of energy and time is given by
\begin{equation}
	\dfrac{\d^2N_i}{\d t\d E} = \dfrac{1}{2\pi}\sum_{\rm dof} \dfrac{\Gamma_i(E,M,a^*)}{e^{E^\prime/T}\pm 1}\,, \label{eq:hawking}
\end{equation}
where $E^\prime \equiv E - m\Omega$ is the total energy of the particle taking into account the BH horizon rotation velocity $\Omega \equiv a^*/(2r_+)$, $a^* \equiv a/M \in [0,1]$ is the reduced spin parameter, $m$ is the projection of the particle angular momentum $l$ and the sum is over the degrees of freedom (dof) of the particle (color and helicity multiplicities). The $\pm$ signs are for fermions and bosons, respectively. The greybody factor $\Gamma_i(E,M,a^*)$ encodes the probability that a Hawking particle evades the gravitational well of the BH.

This emission can be integrated over all energies to obtain equations for the evolution of both PBH mass and spin \cite{PageII1976}
\begin{equation}
	\dfrac{\d M}{\d t} = -\dfrac{f(M,a^*)}{M^2}\,, \label{eq:diffM}
\end{equation}
and
\begin{equation}
	\dfrac{\d a^*}{\d t} = \dfrac{a^*(2f(M,a^*) - g(M,a^*))}{M^3}\,, \label{eq:diffa}
\end{equation}
where
\begin{align}
	f(M,a^*) &\equiv -M^2 \dfrac{\d M}{\d t}\label{eq:fM} \\
	&= M^2\int_{0}^{+\infty} \sum_{\rm dof} \dfrac{E}{2\pi}\dfrac{\Gamma(E,M,a^*)}{e^{E^\prime/T}\pm 1} \d E \,, \nonumber 
\end{align}
\begin{align}
	g(M,a^*) &\equiv -\dfrac{M}{a^*} \dfrac{\d J}{\d t}\label{eq:gM} \\ 
	&= \dfrac{M}{a^*}\int_{0}^{+\infty} \sum_{\rm dof}\dfrac{m}{2\pi} \dfrac{\Gamma(E,M,a^*)}{e^{E^\prime/T}\pm 1}\d E \,. \nonumber 
\end{align}
There are two main effects coming from the PBH spin that play a role in the IGRB. Firstly, a Kerr PBH with a near-extremal spin $a^* \lesssim 1$ radiates more photons than a Schwarzschild one ($a^* = 0$). This is due to the coupling between the PBH rotation and the particle angular momentum for high-spin particles \cite{Chandra4}. We thus expect the constraints to be more stringent. Secondly, a near-extremal Kerr PBH will evaporate faster than a Schwarzschild PBH with the same initial mass due to this enhanced HR \cite{Taylor1998}. Hence, we expect that the constraints will be shifted toward higher PBH masses when the reduced spin parameter $a^*$ increases.

\section{\label{sec:IGRB}Isotropic Gamma Ray Background}

\begin{figure}
	\centering{
		\includegraphics[width = 0.45\textwidth]{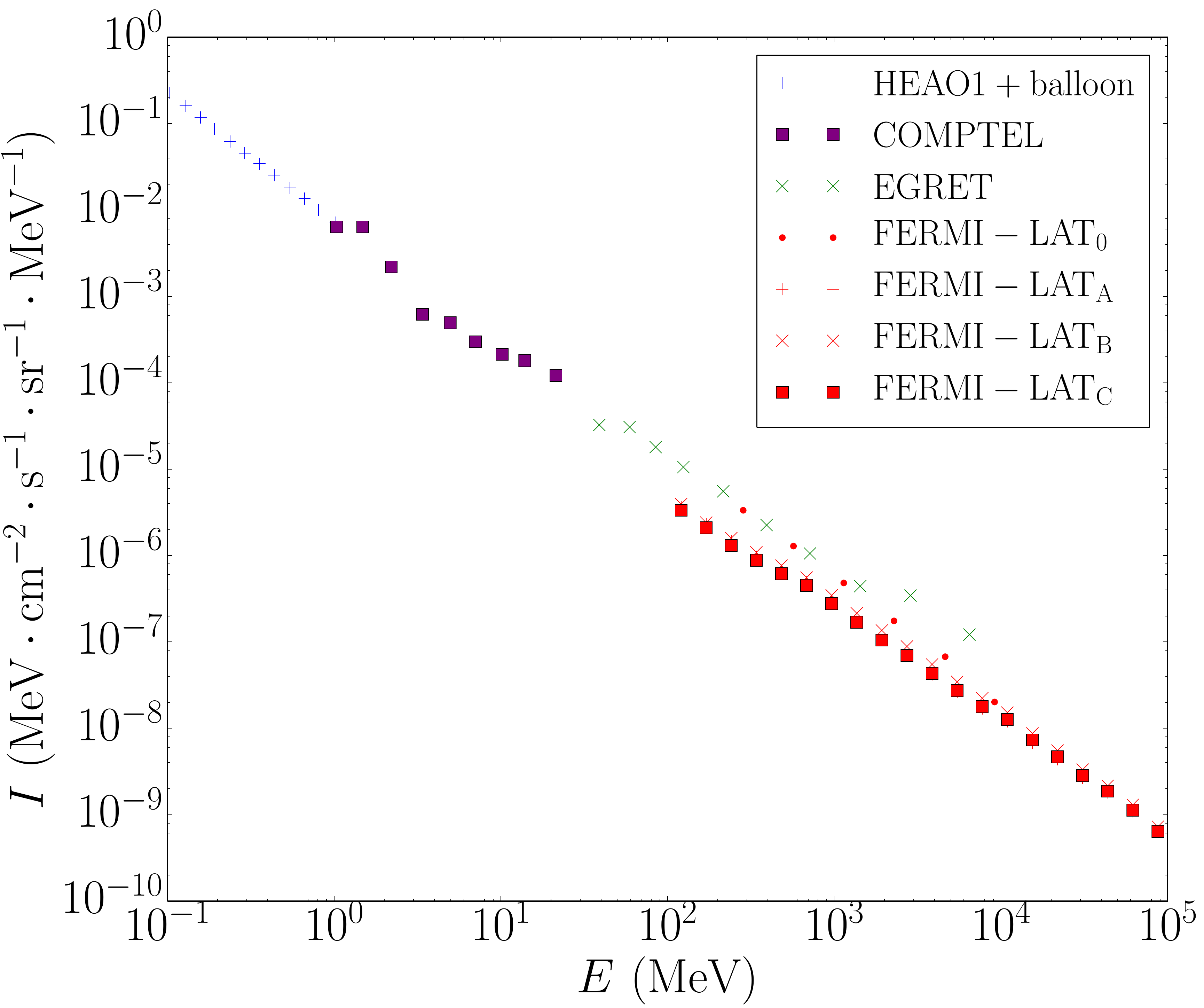}
		\caption{The IGRB as measured by HEAO1+balloon, COMPTEL, EGRET and FERMI-LAT missions \cite{Carr2010,Ackermann2015}. The FERMI-LAT$_0$ marks correspond to the 1st year results and the FERMI-LAT$_{\rm A,B,C}$ marks to 6-year measurements.\label{fig:data_IGRB}}
	}
\end{figure}

Many objects in the Universe produce gamma rays, such as Active Galactic Nuclei (AGN) and gamma ray bursts \cite{Ackermann2015}. The IGRB is the diffuse radiation that fills the intergalactic medium once all point-sources have been identified and removed from the measured photon flux. This background might come from unresolved sources, or more speculatively from DM decays or annihilations. \citefig{fig:data_IGRB} shows the IGRB measured by four experiments (HEAO1+balloon, COMPTEL, EGRET and FERMI-LAT) over a wide range of energies between 100 keV and 820 GeV.

If we consider the simplifying hypothesis that DM is distributed isotropically at sufficiently large scales, then its annihilations/decays should produce, at each epoch of the Universe since transparency, an isotropic flux of photons. Thus, the flux measured along some line of sight should be the redshifted sum over all epoch emissions. Following Carr {\it et al.} \cite{Carr2010}, we estimate the flux at energy $E$ to be
\begin{align}
	I &\equiv E \dfrac{\d F}{\d E} \label{eq:flux_IGRB} \\
	 &\approx \dfrac{1}{4\pi} n\mrm{BH}(t_0) E \int_{t\mrm{min}}^{t\mrm{max}} (1+z(t)) \dfrac{\d^2 N}{\d t\d E}((1+z(t))E)\d t\,,  \nonumber
\end{align}
where $n\mrm{PBH}(t_0)$ is the number density of PBHs of a given mass $M$ today, $z(t)$ is the redshift and the time integral runs from $t\mrm{min} = 380\,000\,$years  at last scattering of the CMB to $t\mrm{max} = {\rm Max}(\tau(M),t_0)$ where $\tau(M)\sim M^3$ is the PBH lifetime and $t_0$ is the age of the Universe. As the Universe is expanding, the number density of PBHs evolves as $(1+z(t))^{-3}$, and the energy of the emitted photons evolves as $(1+z(t))^{-1}$. A last factor $(1+z(t))$ comes from the change of integrand variable from the line of sight to the present time.

\section{\label{sec:results}Results}

We have used the new public code \texttt{BlackHawk} \cite{BlackHawk} to compute the HR of \citeeq{eq:hawking} and the PBH evolution given by Eqs.~\eqref{eq:diffM} and \eqref{eq:diffa}. 
We consider monochromatic PBH distributions 
of masses comprised between $M\mrm{min} = 10^{13}\,$g and $M\mrm{max} = 10^{18}\,$g and initial spin parameters  between $a_{i{\rm,min}}^* = 0$ and $a_{i{\rm,max}}^* = 0.9999$, and  compute the integral of \citeeq{eq:flux_IGRB} over  the redshift (matter-dominated era)
\begin{equation}
	z(t) = \left( \dfrac{1}{H_0 t} \right)^{2/3} -1\,, \label{eq:redshift}
\end{equation}
where $H_0$ is the present Hubble parameter. We then compare the result of the integral to the measured IGRB and find the maximum allowed value of the present PBH number density $n\mrm{PBH}(t_0)$ at a given PBH mass $M$, with a conservative approach taking into account the most stringent constraints (e.g. FERMI-LAT$_{\rm C}$ at $E = 1\,$GeV). The corresponding limit on the DM fraction $f$ constituted of PBHs of mass $M$ is obtained through
$ n\mrm{PBH}(t_0) = f{\rho\mrm{DM}}/{M}\,, 
$
where $\rho\mrm{DM} \approx 0.264\times\rho\mrm{tot} \approx 2.65\times 10^{-30}\,$g$\cdot$cm$^{-3}$ is the current average DM density in the Universe \cite{Planck2018}. If the maximum allowed fraction $f$ is greater than 1, we set it to $1$ in order not to exceed the observed DM density, meaning that the IGRB does not constrain $f$ for the given PBH mass.

\subsection{Monochromatic PBH distribution}

\begin{figure}
    \centering
    \includegraphics[width = 0.45\textwidth]{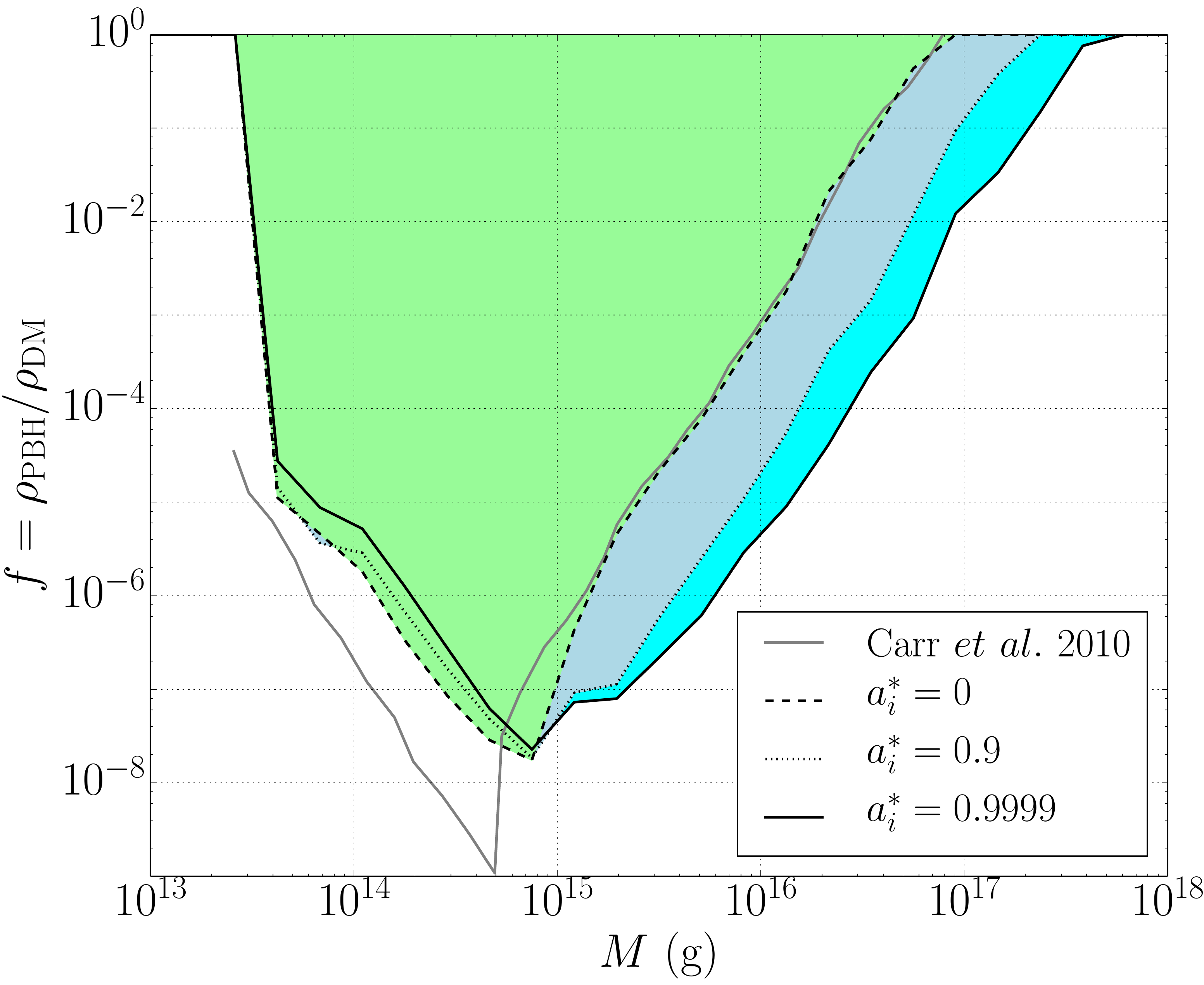}
    \caption{The new IGRB constraints on the DM fraction $f$ in form of PBHs, for monochromatic distributions of PBHs of mass $M_*$ and initial spins $a^*_i \in \{0,0.9,0.9999\}$. Color shaded regions are excluded. For comparison, the result of Carr {\it et al.} \cite{Carr2010} ($a^*_i = 0$) has been superimposed as a gray line.}
    \label{fig:results_mono}
\end{figure}

\citefig{fig:results_mono} shows the resulting constraints for the DM fraction $f$ in PBHs of mass $M_*$ for initial spins $a^*_i \in \{0,0.9,0.9999\}$. First, we see that the $a^*_i = 0$ constraints are comparable with those of \cite{Carr2010}. Our results do not present the feature just after the peak linked to  primary/secondary photons domination explained in this article because we  compute the secondary spectrum for all PBH masses. As a consequence, the peak is smoothed out. We see the second effect anticipated in Section~\ref{sec:Hawking}, that is to say the shifting of the constraint toward higher masses as the initial PBH spin parameter $a_i^*$ increases. This is due to the fact that Kerr PBHs with high initial spin evaporate faster. Thus, in order to have the same kind of HR time-distribution as a Schwarzschild PBH, the PBH must have a higher initial mass. However, this is not accompanied with a more stringent constraint linked to the enhanced emission for Kerr PBHs. We understand  this as follows: PBHs with a higher mass emit photons at lower energies (cf. the temperature-mass relation \citeeq{eq:temperature}) where the IGRB constraints are less severe. The two effects approximately cancel. The main result that we find is that if PBHs have a high initial spin parameter $a^*\lesssim 1$, the ``small-mass" window $10^{17}-10^{19}\,$g can be closed up to almost one order of magnitude on its lower boundary. For the possible existence of such high spins, see for example \cite{ExtremalSpins}.

\subsection{Extended PBH distribution}
\label{sect:extended}

We also obtained constraints for extended mass functions to study the effects related to the width of the peak. Some pioneering work has been done in \cite{Carr2017,Kuhnel2017,Bellomo2018,Lehmann2018} concerning extended mass functions, predicting that the constraints on an extended distribution should be more stringent than the expected constraint resulting from the addition of monochromatic distributions. The conclusion of these papers is that a simple conversion from monochromatic to extended mass functions is not analytically trivial. We thus derive the extended mass function constraints by computing the full Hawking spectra associated to them before applying the constraints.

\begin{figure}
	\centering{
		\includegraphics[width = 0.45\textwidth]{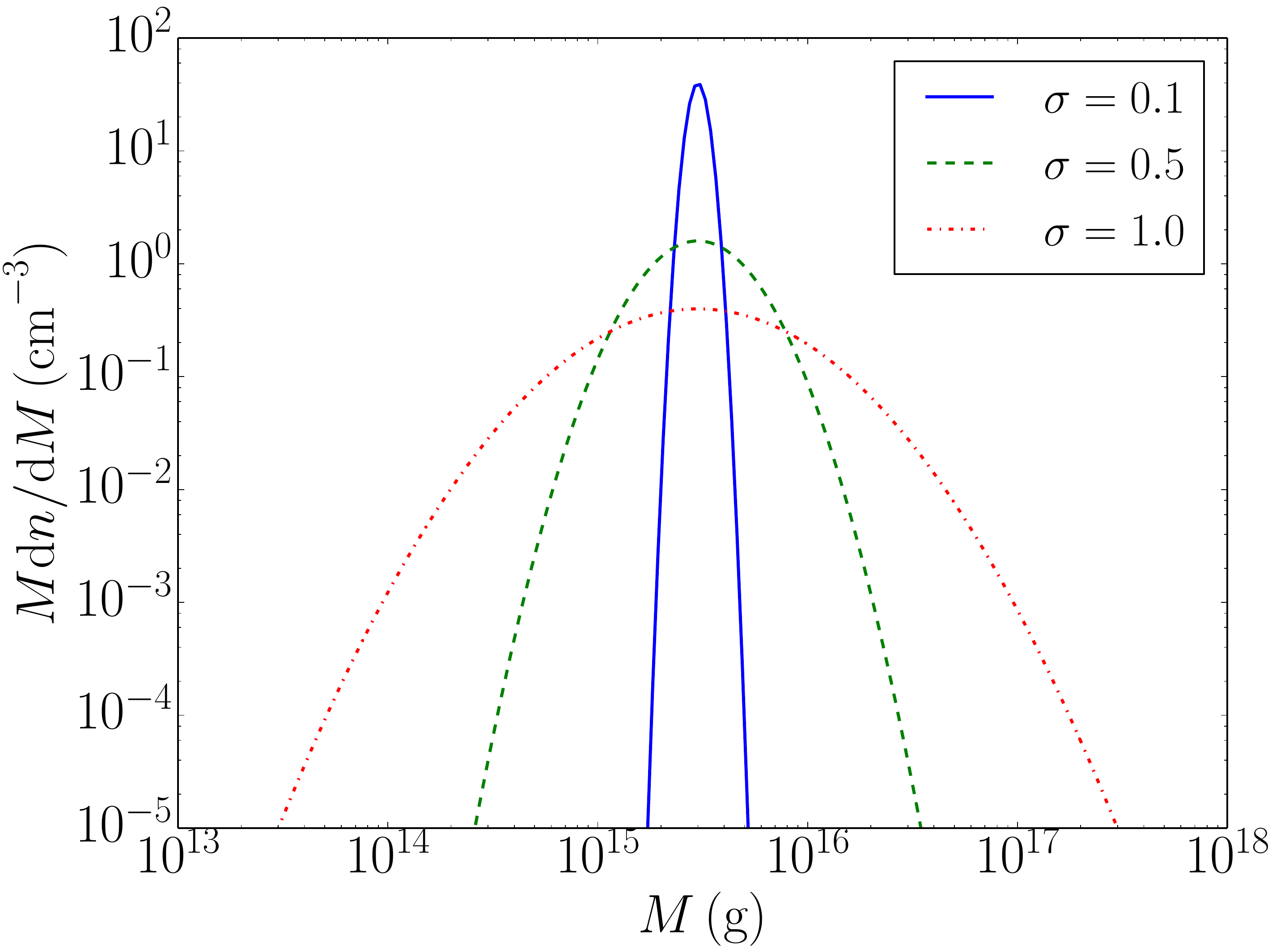}
		\caption{Examples of distributions following \citeeq{eq:ext_dis} for values of $\sigma \in \{0.1,0.5,1\}$. The amplitude is $A = 1$ and the central mass is $M_* = 3\times10^{15}\,$g for all distributions for clarity.\label{fig:ext_dis}}
	}
\end{figure}

We considered extended mass functions of log-normal form
\begin{equation}
	\dfrac{\d n}{\d M} = \frac{A}{\sqrt{2\pi}\sigma M}\exp\left( -\dfrac{(\ln(M/M_*))^2}{2\sigma^2} \right)\,, \label{eq:ext_dis}
\end{equation}
{\it i.e.} a Gaussian distribution in logarithmic scale for the density. $A$ is some amplitude, linked to the fraction of DM into PBHs. This distribution is normalized for $A = 1$. To compute the spectra, \texttt{BlackHawk\_tot} \cite{BlackHawk} was used with $\texttt{spectrum\_choice} = 5$, and 10 different PBH masses scanning the whole peak width.

We do not assume any model of PBH formation to justify this distribution, which is based on the fact that a Gaussian peak can mimic any peak in the PBH distribution resulting from a particular mechanism of formation, but we note that this mass distribution -- with some variations -- has been used in various works linked to different PBH formation mechanisms and mass ranges \cite{Green2016,Kannike2017,Calcino2018,Boudaud2019,DeRocco2019,Laha2019}. We have done a similar scan to the one described in the previous section, with $M_*$ the mean of the Gaussian distribution ranging from $10^{13}\,$g to $10^{18}\,$g (cf. the Introduction for the PBH mass bounds), and its width $\sigma \in \{0.1, 0.5, 1\}$. \citefig{fig:ext_dis} shows examples of these distributions for $M_* = 3\times10^{15}\,$g.

\begin{figure*}
	\centering{
		\includegraphics[width = 1.\textwidth]{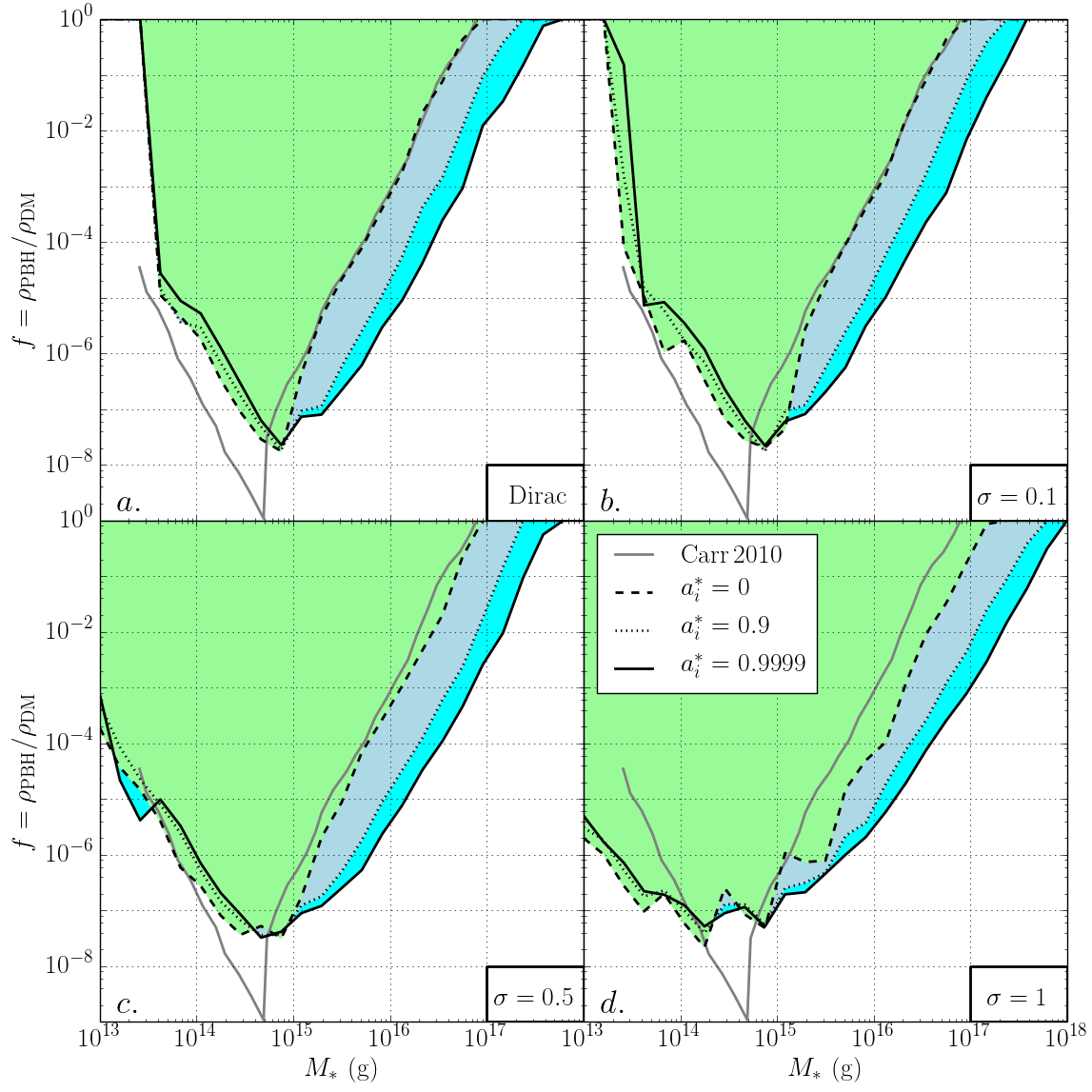}
		\caption{$a.$ -- The same monochromatic plot as \citefig{fig:results_mono} for comparison (here $M_*$ is the monochromatic mass).\\
		$b.,\, c.,\, d.$ -- The IGRB constraints on the DM fraction $f$ in form of PBHs, for distributions of PBHs of initial spins $a^*_i \in \{0,0.9,0.9999\}$ following \citeeq{eq:ext_dis}, with central mass $M_*$ and widths $\sigma \in \{0.1,0.5,1\}$ ($b.,\,c.,\,d.$ respectively). Color shaded regions are excluded.\label{fig:results}}
	}
\end{figure*}

\citeeq{eq:flux_IGRB} must be modified to obtain the fraction for an extended mass function. The flux is now given by
\begin{align}
	I &\approx \dfrac{1}{4\pi} E \int_{t\mrm{min}}^{t\mrm{max}} (1+z(t)) \dfrac{\d^2 n}{\d t\d E}((1+z(t))E)\d t \label{eq:flux_IGRB_ext} \\
	 &\approx \dfrac{1}{4\pi} E \int_{t\mrm{min}}^{t\mrm{max}} (1+z(t)) \nonumber \\ &\times \int_{M\mrm{min}}^{M\mrm{max}}\left[\dfrac{\d n}{\d M}\dfrac{\d^2 N}{\d t\d E}(M,(1+z(t))E)\,\d M\right]\d t\,.  \nonumber
\end{align}
with $\d n/\d M$ given by \citeeq{eq:ext_dis}. The fraction of DM in form of PBHs is obtained by maximizing this flux (increasing the normalization constant $A$) while respecting all the IGRB constraints, and given by
\begin{align}
    f &\equiv \dfrac{\rho\mrm{PBH}}{\rho\mrm{DM}} \\
    &= \dfrac{A}{\rho\mrm{DM}\sqrt{2\pi}\sigma} \int_{M\mrm{min}}^{M\mrm{max}} \exp\left( -\dfrac{\log(M/M_*)^2}{2\sigma^2} \right)\d M\,.
\end{align}
It is again limited to 1 in order not to exceed the DM content of the universe. Even if the IGRB constraints valid at $M_*$  prevent $A$ from exceeding its maximum value when $\sigma \rightarrow 0$ (monochromatic distribution), we expect that when the distribution width $\sigma$ increases, monochromatic IGRB constraints from $M \lesssim M_*$ and $M\gtrsim M_*$ will become more and more important, thus limiting $A$. On the other hand, if $\sigma$ increases, the full distribution integral that contributes to the DM fraction $f$ increases as well because of the $M\lesssim M_*$ and $M\gtrsim M_*$ contributions. The competition between the two effects is difficult to forecast.


\citefig{fig:results} (panels $b$, $c$ and $d$) shows the constraints for distribution widths $\sigma \in \{0.1,0.5,1\}$ (respectively) and $a^*_i \in \{0,0.9,0.9999\}$. There are 3 kinds of observations to be considered.

1) For a fixed PBH initial spin $a^*$, when the width of the distribution $\sigma$ increases, the excluded region widens. This effect is sensible when $\sigma \gtrsim 0.5$. Indeed, for a sharp distribution, the IGRB constraints that play a role in limiting $f$ are those close to the central mass $M_*$. When the distribution gets wider, constraints from masses far from the central mass are important. As the constraints are the most severe for $M\mrm{peak}\lesssim 10^{15}\,$g, wide distributions centered on $M_* \ll M\mrm{peak}$ and $M_* \gg M\mrm{peak}$, which have a tail reaching the peak mass, are severely constrained. This extends the excluded region to $M \ll M\mrm{peak}$ and $M \gg M\mrm{peak}$ and closes the $10^{17} - 10^{18}\,$g window for all DM made of PBHs.

2) For a fixed PBH initial spin $a^*$, when the width of the distribution $\sigma$ increases, the constraint on $f$ close to the peak $M\mrm{peak}$ decreases. This is due to the fact that the amplitude $A$ of the mass distribution is most severely constrained by the $M\approx M\mrm{peak}$ contribution. If we extend the mass distribution around $M\mrm{peak}$, we do not add new strong IGRB constraints, but we increase the mass fraction $f$ of DM into PBHs.

3) For a fixed width of the distribution $\sigma$, when the initial spin $a^*$ of the PBHs increases, the constraints are shifted toward higher central masses while being slightly more stringent. This is coherent with the results of \citefig{fig:results} (panel $a$) for the monochromatic distributions.

We can sum up these observations in the following way. For an extended PBH mass function, the overall constraint comes from the PBHs evaporating  today in this distribution with initial mass $M = M\mrm{peak}$. Distributions centered away from $M\mrm{peak}$ are more and more constrained as the tail of the distribution is important at $M\mrm{peak}$: $f$ decreases as $\sigma$ increases because the maximum value of $A$ decreases. Distributions centered close to $M\mrm{peak}$ are not much more constrained when the distribution expands, the maximum value of $A$ remains quite the same: $f$ increases as $\sigma$ increases because the distribution integral increases. The very same effects can be observed on the right panel of Fig.~2 of \cite{Boudaud2019}.

\subsection{Comparison to other constraints}

Recent works have tried to close the very same mass range for PBHs constituting all of the dark matter, showing that this scenario is attracting much attention. Future femtolensing \cite{Katz2018} or X-ray \cite{Ballestros2019} surveys as well as galactic positrons data \cite{Boudaud2019,DeRocco2019,Laha2019} all constrain the same $M = 10^{16} - 10^{18}\,$g mass range.

\begin{figure*}
    \centering
    \includegraphics[width = \textwidth]{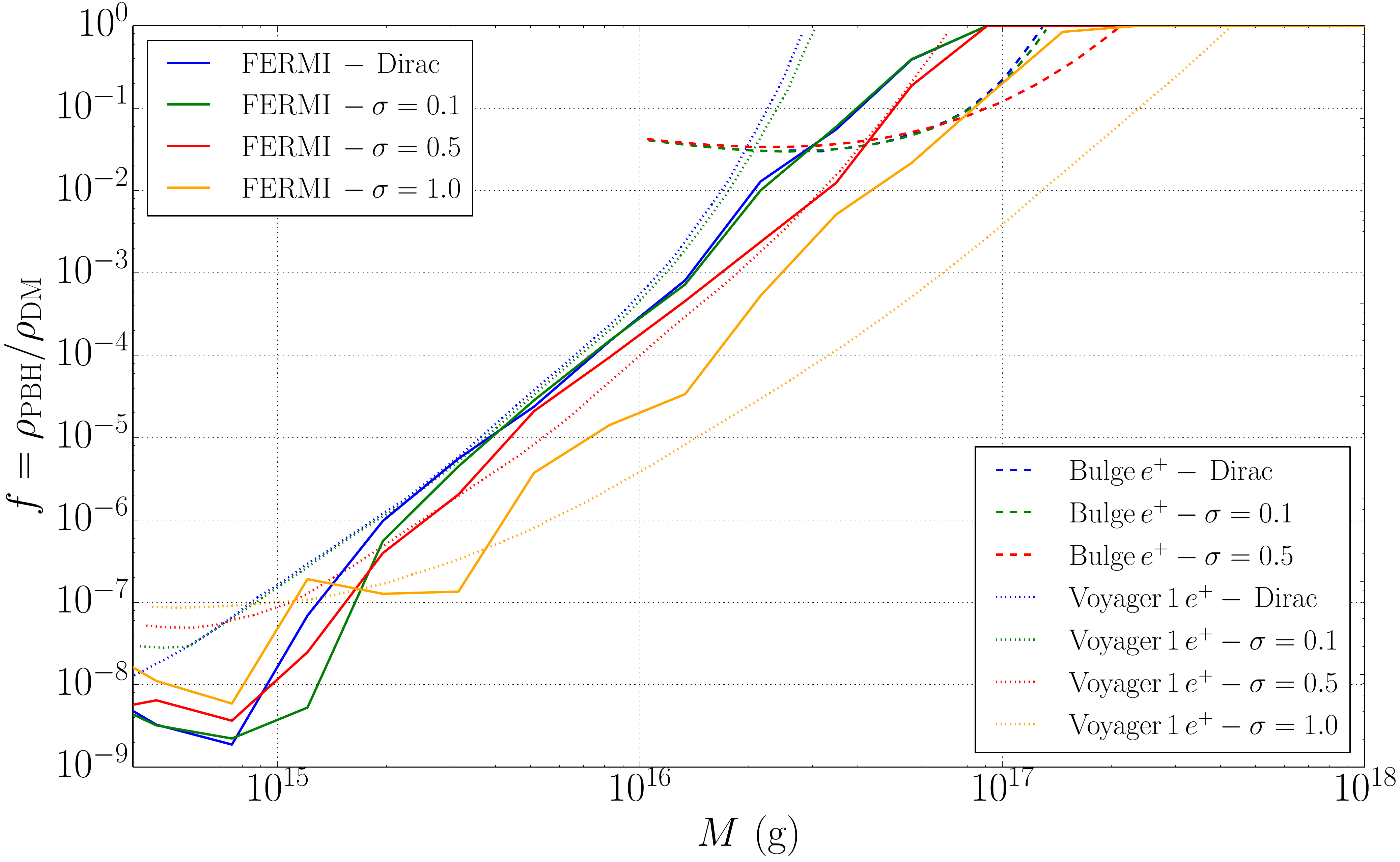}
    \caption{Comparison between the limits derived in this work (solid lines) and those coming from electron-positron annihilating in the Milky Way bulge (dashed lines) \cite{DeRocco2019} and the local Voyager 1 positron detections (dotted lines) \cite{Boudaud2019}. Limits are shown for monochromatic mass functions (blue) as well as log-normal mass functions with gaussian width $\sigma \in \{ 0.1,0.1,1\}$ (green, red and orange respectively). Ref.~\cite{DeRocco2019} does not provide the $\sigma = 1$ data.}
    \label{fig:comparison}
\end{figure*}

\citefig{fig:comparison} compares the limits derived here and those based on galactic positrons \cite{Boudaud2019,DeRocco2019} (we do not include the extended mass function limits of \cite{Laha2019} because they result from a convolution of monochromatic limits with the mass function, a method which was decried -- see Section~\ref{sect:extended}). The limits obtained with a local measurement of the positron flux by Voyager 1, which has recently leaved the heliosphere and is capable of detecting low-energy positrons, are somewhat of the same order of magnitude as ours for widths $\sigma \lesssim 0.5$, while becoming significantly more stringent for $\sigma \gtrsim 1$. The limits derived from electron-positron annihilation in the galactic bulge -- thus contributing to the $511\,$keV photon line -- are more severe than the 2 others in the central mass region $M \sim 10^{17}\,$g, but the authors claim that the Voyager limits are more restrictive when $\sigma \gtrsim 1$. As those limits come from totally different galactic and extra-galactic measurements, we consider them as most interestingly independent and complementary, increasing the robustness of the conclusion concerning PBHs not constituting all of the DM in the $10^{16} - 10^{17}\,$g mass range.

\section{Conclusion}

In this Letter, we have updated the IGRB constraint on PBH evaporation for monochromatic Schwarzschild PBH distributions, using the latest FERMI-LAT data and the new code \texttt{BlackHawk}. This has resulted in enhancing the constraint on the masses of presently evaporating PBHs, and reducing the constraint on $M\mrm{peak} \lesssim 10^{15}\,$g.
Our main result is the extension of the IGRB constraint from Schwarzschild to Kerr PBHs, and from monochromatic to extended mass functions. We have shown that increasing the initial spin parameter $a_i^*$ of PBHs to near extremal values can close the mass window $10^{17} - 10^{19}\,$g (where PBHs could still represent all of the DM). We have also demonstrated that extended mass functions can allow a greater fraction of DM in the form of PBHs when they are centered close to the strongest monochromatic constraint, while they are more severely constrained when centered away from this peak. In this case,  the allowed mass window can be reduced even with Schwarzschild PBHs, complementing previous work in the same mass range with positron emission by evaporating PBHs.

\section*{Acknowledgments}

We would like to thank P.~Graham and W.~DeRocco for useful private discussion and for pointing out an error (corrected).

\bibliographystyle{apsrev4-1}
\bibliography{biblio}

\end{document}